# SOFI for Plasmonics: Extracting Near-field Intensity in the Far-Field at High Density

Robert C. Boutelle[1], Xiyu Yi[1], Daniel Neuhauser[1,2], Shimon Weiss[1,2,3,4]

[1]Department of Chemistry & Biochemistry, [2]California NanoSystems Institute, and [3]Department of Physiology, University of California Los Angeles, Los Angeles, California 90095, United States
[4] Department of Physics, Bar Ilan University, Ramat Gan, 52900, Israel

**Introduction**

Super-resolution microscopy is the classification of techniques that allow imaging below the diffraction limit faced by conventional optical microscopy. Over the past decade, super-resolution imaging has rapidly grown and greatly impacted the field of biology, as evidenced by the 2014 Nobel Prize in chemistry[1]. Its rapid growth has stemmed various methods, including the early STORM, PALM, STED and newly developed PAINT, GSDIM, and SOFI[2-7]. Additionally, super-resolution techniques have gained popularity in the field of materials, influencing polymers and catalysis [5, 8-12]. These far-field super-resolution techniques are based on individual emitters that are modulated between emissive ("on") and non-emissive ("off"). From this modulation, one can spatially separate two emitter point spread functions (PSFs) in time, captured in the far-field as a stream of images, such that a single emitter PSF can be fit to a model function, such as a 2D-Gaussian. Superimposing the positions of the individual emitters creates a composite reconstructed image[2, 13, 14]. Recently, there has been a push to utilize super-resolution imaging for plasmonics[15]. However, there are many technical challenges that must be overcome before super-resolution for plasmonics becomes feasible.

Plasmonics and super-resolution imaging naturally compliment each other. Plasmonics provides a simple strategy to achieve subdiffraction excitation volumes using a far-field excitation source. Additionally, plasmonic nanostructures are by their very nature nanoscale in dimension, requiring subdiffraction-limited imaging tools to measure the complex relationship between structure, resonance frequencies, and local field enhancement[16]. With the development of plasmonics-based devices and circuits there is a growing need for detecting and characterizing plasmonic effects. At first glance, super-resolution imaging appears to be the best method to characterize plasmonic systems in the far-field. When applied to plasmonic nanostructures, dye emission intensity is typically used as a far-field reporter of the near-field intensity[17]. Due to the strong electric field near the plasmonic nanostructure, the dye has enhanced emission when the excitation or emission frequency matches the plasmon resonance. Unfortunately there are detrimental problems with probing the near-field intensity with far-field emitters.

First, the diffraction limited emission is affected via a combination of mechanisms, including plasmon-molecule coupling, point spread function distortions, and image dipole formation[15]. These mechanisms lead to erroneous error in localization. Second, fluorescence enhancement changes non-monotonically as a function of distance from the metallic surface. When the emitter is too close to the surface (< 10 nm), the fluorescent rate drops sharply due to quenching effects[17]. Thus fluorescent emission intensity is not a good candidate for measuring plasmon near-field intensity in close proximity to the metallic surface. In this report, we focus on a new method to measure the near-field plasmonic fields using a far-field reporter that is invariant to quenching effects. Methods to mitigate the PSF distortion and error in localization are beyond the scope of this work but are outlined by others [15]. Future reports will integrate methods to allow more accurate characterization of the plasmonic near-field.

Specifically, we previously created a method to measure plasmonic near-field intensities dubbed **C**haracterizing **O**ptical **F**ield **I**ntensity by **Bl**Inking **N**anoparticle**S** (COFIBINS). [18] This method extracts plasmonic near-field intensity in the far-field using QD stochastic blinking field sensitivity. COFIBINS demonstrated that in low areal density (< 1.8 QDs/µm$^2$), individual QD time-intensity trajectory could be used to measure the change in blinking rate as a measure of optical field intensity. COFIBINS was applied to silver nanowires (AgNWs) and accurately extracted the propagation length and penetration depth of the plasmonic waveguides. The downside is that COFIBINS had low spatial resolution. The QDs had to be well separated to obtain their individual time-intensity trajectory for accurate characterization of the plasmonic near-field.

Recently, we developed a method to extract photo-physical properties of QDs at high labeling densities. A variant of superresolution optical fluctuation imaging (SOFI) methodology [4, 7, 19-21] is used to spatially resolve features at high density and, in a method similar to bSOFI [22] that utilizes multiple orders of cumulant analysis (dubbed as MOCA-SOFI), extract the photo-physical information of QDs at high labeling density. When combined with COFIBINS, this method creates a crucial development for true far-field super-resolution detection of plasmonic near-fields.

A direct demonstrate of high spatial resolution measurement of SOFI-COFIBINS on plasmonic near-field intensity in the far-field is done here on silver nanowires (Ag NWs). Ag NW waveguides have an inhomogeneous (decaying) field distribution both in the surface plasmon polariton (SPP) propagation direction (along the long direction of the wire) and penetration depth direction (perpendicular to the long direction of the wire). SOFI is used in combination COFIBINS to extract the QD blinking parameter in high areal densities along these Ag NWs. QDs were spin-coated at high areal density (>30 QDs/µm) on top of Ag NWs. These experiments test the efficacy of the newly developed SOFI-COFIBINS method. The method extracts values comparable to the fluorescence intensity yet contains super-resolved information.

## Results and Discussion
### High Density QD Photo-Physical Parameter Extraction
Previous studies have already demonstrated that QD blinking is relatively invariant to enhancement and quenching effects, yet their blinking rate is sensitive to optical near-field intensity. Optical field intensities affect the stochastic switching of QDs between an "on" state with a high photon emission rate and an "off" low emission state. Stochastic switching, also known as "intermittency", or "blinking", has been studied extensively, both experimentally and theoretically. Auger recombination is commonly invoked to explain blinking, but other processes, such as surface and heterointerface charge trapping, have also been shown to contribute to the switching. Auger recombination is a three-particle process that results in a nonradiative transition due to the absorption of energy from an exciton by a spectator particle, leaving the QD charged, and in a "dark", or "off", state. Only once the charged QD is neutralized does the emissive "on" state resume. Blinking in QDs is inherently stochastic and independent of other nearby QD emitters. Traditionally, emission from a single QD is recorded in time bins to analyze the blinking phenomena. A threshold is used to discern "on" and "off" time periods in the telegraph noise-like time trajectory and histograms are constructed for the "on" and "off" periods. At low excitation power, the histograms exhibit a near perfect power law distribution for both the "on (+)" and "off (−)" states, described by $P_\pm(t) = t^{-m}$. For higher excitation energies, the "on" time distribution starts to bend at long "on" times ($P_+ = t^{-m}e^{-\Gamma t}$, where $\Gamma$ is the intensity bending parameter). Here, m is the slope of the on/off-time probability distribution in a log–log plot. Thus, the stronger the excitation intensity, the shorter the "on" periods and the more likely the QD will be to enter an "off" state. The bending parameter, $\Gamma$, yields information on the excitation field via the blinking statistics of individual QDs.

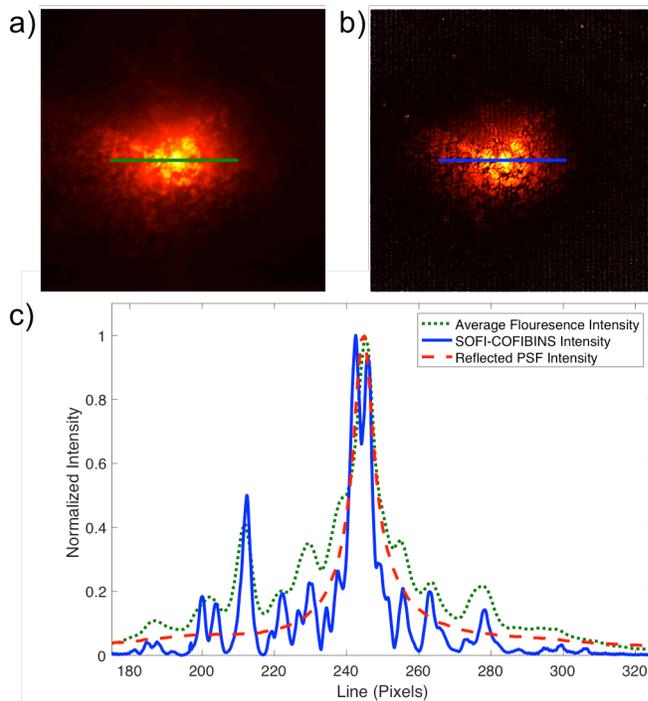

**Figure 1:** Extracting intensity values from a defocused PSF. **(a)** Average fluorescence intensity per pixel. **(b)** Extracted $\Gamma$ value per pixel via SOFI-COFIBINS. **(c)** Cross section of **(a)** (green) and **(b)** (blue) plotted against the reflected PSF Intensity. SOFI-COFIBINS agrees well with both the average fluorescence intensity and reflected PSF intensity.

The autocorrelation function (ACF) of a QD's individual time-intensity trajectory can also extract the optical intensity bending parameter $\Gamma$ by fitting:

$$1 - G(\tau) = \beta(m)\theta^{m-1}\Gamma^{2-m}\tau^{2-m}$$

where $\theta$ is the minimum capture window and $\beta(m)$ is a numerical function (product of $\Gamma$-functions) of $m$. The QD blinking parameter $\Gamma$ shows a linear trend for power dependent parameter over three orders of magnitude optical power density. Thus, by taking the autocorrelation of the time-intensity trajectory of a blinking QD, the optical field strength can be measured. This method has successfully been applied to Ag NW waveguides. The largest drawback is the inability to resolve the individual time-intensity trajectory at high densities. The solution is to use SOFI to extract out the blinking parameters and simultaneously super-resolve the image at high density.

In SOFI, the temporal autocorrelation amplitude of pixels with signal contribution from multiple QDs is lower than the autocorrelation amplitude of pixels with signal contribution from only one QD, yielding an image with a dip between QDs. This dip leads to shrinkage of the individual PSF as a function of $1/\sqrt{n}$, where $n$ is the SOFI order, and is the basis for increased resolution. We have taken this method a step further to extract out the photophysical properties of QDs at high density. Briefly, the fluctuation signal within a given pixel will depend on the emitter PSF, emitter brightness, and its fluctuation profile, which contains an "on-"/"off-time" ratio, such that:

$$F(r,t) = \sum_k U(r - r_k) \cdot \epsilon_k \cdot S_k(t)$$

where $F(r,t)$ is the signal acquired in time ($t$) on a single pixel ($r$), $k$ is the emitter index, $U(r - r_k)$ is the model PSF (in this case 2D-Gaussian), $\epsilon_k$ is the emitter brightness, and $S_k(t)$ is the fluctuation profile. Then, in MOCA-SOFI we extract out the individual parameters. Details on the method is beyond the scope of this report and will be presented in a subsequent independent paper.

The key conclusion is that the fluctuation signal in a single pixel containing multiple QD fluctuation profiles can be decoupled and resolved. Specifically, we acquire the "on-"/"off-time" ratio of each quantum dot contributing to the signal of a single pixel such that:

$$\frac{average\ on\ time}{average\ off\ time + average\ on\ time} = \frac{\int_\theta^\infty tP_+(t)dt}{\int_\theta^\infty tP_-(t)dt + \int_\theta^\infty tP_+(t)dt}$$

where $P_+(t)$ and $P_-(t)$ are the probability distribution function of the on/off lifetimes respectively:

$$P_+(t) = At^{-m}e^{-\Gamma t}$$

$$P_-(t) = Bt^{-m}$$

where $A$ and $B$ are the normalization constant, $m$ is the power law slope, and $\Gamma$ is the excitation intensity bending parameter. Combining equations [EQUATIONS], we get:

$$\frac{\Gamma^{2-m}\beta}{\frac{\theta^{2-m}}{2-m} + \Gamma^{2-m}\beta}$$

where $\beta$ is a place holder for the gamma function $\Gamma(2 - m)$. Thus, we can extract out the $\Gamma$ intensity parameter and, when combined with COFIBINS, map with super-resolution the optical field intensity.

A proof of principle demonstration is illustrated in Figure 1 with a defocused laser PSF. 800nm emission QDs were deposited via spincoat on a coverslip in high areal density. Next a thick layer of PMMA was deposited via spincoat to protect the QDs. The QDs were then illuminated with a defocused 642nm CW laser. An EMCCD camera captured the far-field emission time-trajectories using a conventional diffraction-limited wide-field optics and a 750 long pass filter (Edmund optics). 20K frames with exposure of 30ms were captured for analysis.

An image constructed from fitted bending parameters ($\Gamma$) for each pixel maps the localized electric field and super-resolve the defocused PSF illumination. We can see that the SOFI-COFIBINS result agrees well with both the average fluorescence intensity and the reflected PSF intensity. This result illustrates a proof of principle that the optical field strength can be extracted at high density through blinking statistics with SOFI-COFIBINS.

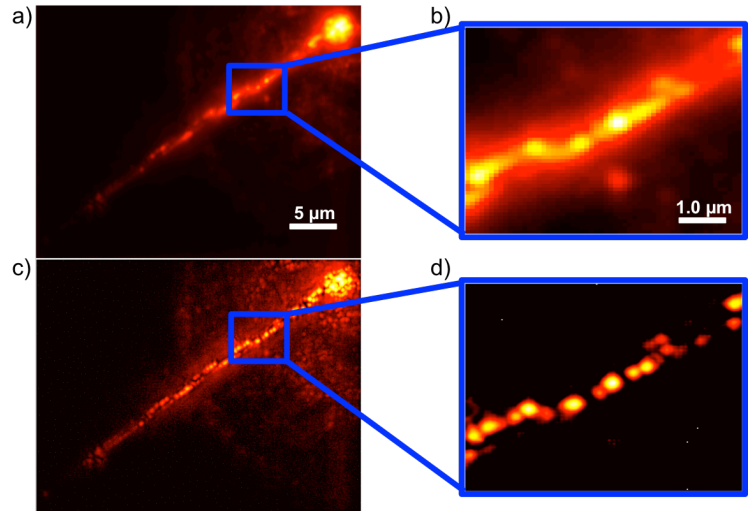

**Figure 2: Superresolved Plasmonics** QDs were drop-casted on glass converslip containing randomly oriented AgNWs. SPPs were excited with a confocal 642nm CW laser at one end of AgNWs (top right of each image). The SPPs evanescent field excites the QDs downstream the wire. **(a)** Average pixel intensity. **(b)** Zoom-in to boxed area in (a). **(c)** QD on-off ratio extraction via SOFI-COFIBINS. **(d)** Zoom-in to boxed area in (c) All heat-maps are log scale (to allow visualization of the plasmon propagation exponential decay).

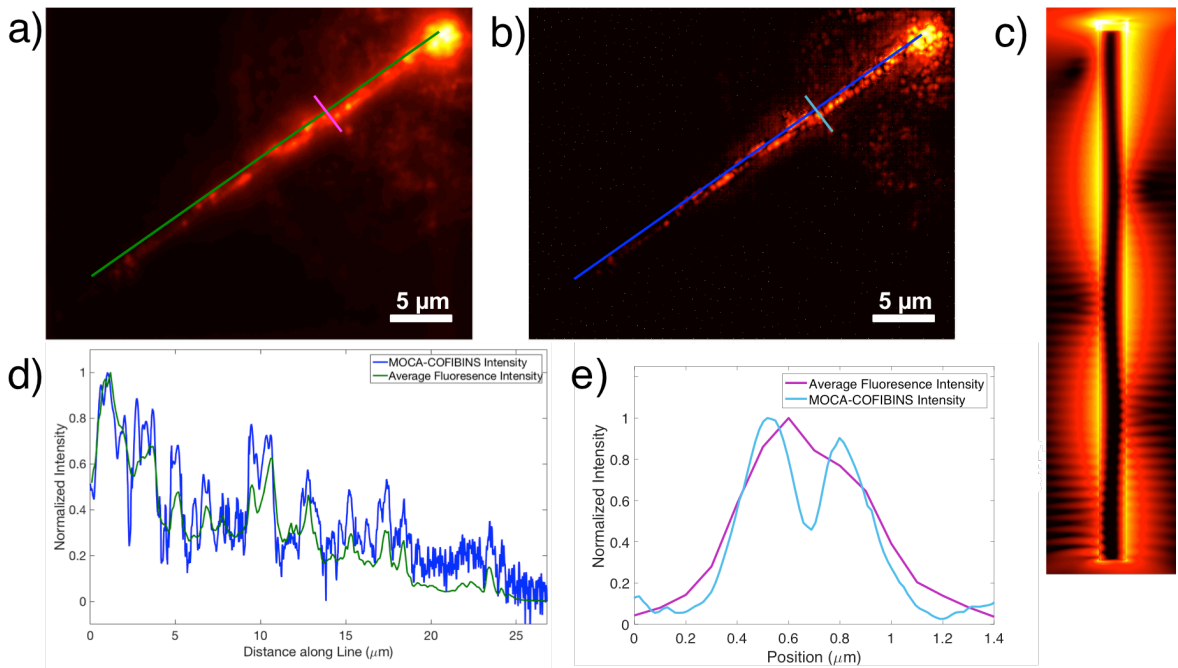

**Figure 3: Cross Section analysis of SOFI-COFIBINS (a)** Average pixel intensity. **(b)** QD on-off ratio extraction via SOFI-COFIBINS. **(c)** FDTD simulations illustrating the modes visualized are due to a 45° polarization angle of excitation. **(d)** Normalized cross section for SPP propagation of average pixel intensity from (a) (green) and SOFI-COFIBINS (b) (blue) plotted together. **(e)** Normalized cross section for SPP penetration of average pixel intensity from (a) (magenta) and SOFI-COFIBINS (b) (teal) plotted together. The two show similar trends yet the SOFI-COFIBINS has super-resolved information.

**SOFI-COFIBINS for Plasmonics**

A direct application of our merged SOFI-COFIBINS to the characterization of plasmonic waveguide is demonstrated next. QDs were drop-casted at high concentration onto a 220nm AgNW. Upon confocal excitation of the SPPs at one end of the wire (642nm CW laser), the QDs excite and emit (800nm) downstream due to the SPP's evanescent field. Figure 2a shows the average (sum image) of 10k acquired frames. Figure 2c shows the superresolved image using SOFI-COFIBINS. The two methods give similar trends to the plasmonic SPP. However, SOFI-COFIBINS has superresolved information.

This is further verified by taking cross sections for the propagation and penetration direction of the SPP (Figure 3). Here the average fluorescence intensity again shows similar trend to the SOFI-COFIBINS super-resolved image to extract the SPP propagation length and penetration depth.

## **Conclusion**

Our initial implementation of SOFI-COFIBINS shows promise for simultaneous field intensity extraction with the $1/\sqrt{n}$ spatial resolution enhancement factor, where $n$ is the SOFI order, typical of SOFI analysis. However, we need to investigate our method's limitations such as large electromagnetic field gradient, time resolution, and dynamic range. Optimization will be made by simple simulation of fake QD emitters with generated intensity trajectories and using known blinking profiles, spread along a plasmonic system of simulation. FDTD simulations will dictate the fields experienced by these fake QDs, and therefore their intensity trajectories and PSF profile. The emission profile, generated as a movie, will be analyzed by SOFI-COFIBINS and the errors in the output compared with the preassigned locations will yield a measure of the method's accuracy. We will use this information to optimize under what conditions SOFI-COFIBINS best performs. The optimized conditions will guide our experimental protocol and what we learn from our experimental measurements will help optimize our theoretical simulations and SOFI-COFIBINS analysis.

## **Contributions**

Robert C. Boutelle was responsible for experimental measurement/analysis, application of MOCA-SOFI, and FDTD simulations. Xiyu Yi was responsible for development and coding of MOCA-SOFI in addition to advising Robert on the SOFI algorithm. Daniel Neuhauser oversaw the theoretical FDTD results. Shimon Weiss managed the project and science.